# A Distributed Parallel Model to Analyze Journal Impact Factors

ZHOU, Jian[1]    CAI, Ning[1, 2 *]    TAN, Zong-Yuan[1]

[1] College of Electrical Engineering, Northwest Minzu University, Lanzhou, China
[2] Key Laboratory of China's National Linguistic Information Technology, Lanzhou, China

**Abstract:** A simple abstract model is developed as a parallel experimental basis for the aim of exploring the differences of journal impact factors, particularly between different disciplines. Our model endeavors to simulate the publication and citation behaviors of the articles in the journals belonging to a similar discipline, in a distributed manner. Based on simulation experiments, the mechanism of influence from several fundamental factors to the trend of impact factor is revealed. These factors include the average review cycle, average number of references and yearly distribution of references. Moreover, satisfactory approximation could possibly be observed between certain actual data and simulation results.

# 1. Introduction

Journal Impact factor (JIF) is the most important index in SCI, which is a quantitative tool for evaluating the ranks and the grades of various scientific journals in the Journal Citation Report (JCR) database. The vast majority of the universities and research institutions rely on impact factor as an effective tool for sorting and assessing the scientific research performance of scientists through their publications [1-2]. Even some publishers consider impact factor values as an indirect marketing reference for selling their journals. Furthermore, although the original purpose of the impact factor is to measure the quality of academic contributions, impact factor has already been used not only in the bibliometric field, but also in many decision tasks increasingly, such as research grants allocation and journal subscriptions. Even in Finland and Spain, journal impact factor has been brought into law, aiming to improve the overall level of sciences of the country [3]. In some sense, impact factor has been perhaps regarded as the most valid means to measure the quality of scientific products.

[1] Corresponding author: Cai, Ning (Email: caining91@tsinghua.org.cn)

With a deeper understanding and application of the impact factor, the role and limitations of the impact factor in the process of scientific evaluation have been addressed by more and more scholars [4-5]. For example, as early as 1978, in order to overcome the subject bias of citation measure, the concept of Disciplinary Impact Factor (DIF) is introduced by Hirst [6]. Moreover, Hemmingsson and Yang *et al.* [7-9] concerned the situation that the impact factor could be artificially manipulated by editors of scientific journals. Garfield [10] pointed that the impact factor should be used with caution, in view of the potential problem caused by that the two-year citation window of the JCR is too short to perceive the real influence of journals in relatively slowly evolving disciplines. It thus can be seen that using impact factor to scientifically measure the quality of scientific journals remains a puzzling question, which can be attributed to the fact that the impact factor of a journal is determined by multi-factors [11-12], or in other words, it is improbable that any single index is suitable for describing the citation of all journals.

In the present paper, we focus on the study of significant qualitative differences of impact factors between different disciplines and endeavor to determine how the impact factor is affected by the fluctuation of certain factors in publication. For instance, as statistical conclusions, our observation indicates distinct monotonous correlations between the value of impact factor and the two fundamental factors, i.e. average number of references and the general range of the yearly distribution of references, respectively. Moreover, the shorter the average review cycle of a journal, the higher the impact factor of the journal tends to be. In comparison with other methods, the key idea of our approach, described in the subsequent section, is based on distributed experimental modeling [13-15] for empirically simulating the trends of impact factors of various types of journals and analyzing the statistical correlations between impact factor and certain publication factors.

The study is mainly based on the approach of social computing, with a combination of both empirical and analytical analysis, in which the empirical analysis here is rooted within the theoretical framework of parallel systems [16-17]. We study the laws of journal impact factor by building and observing the behaviors of virtual simulation systems. The objective of simulation systems is not for comprehensively and quantitatively mimicking the real world, instead, it could be very conducive to drawing conclusions about certain issues, qualitatively, in particular those negative conclusions asserting that something should not happen. We hope our research could provide theoretical hints for deeper understanding of the mechanism of JIF dynamics and facilitating further enhancement in managing academic journals.

This paper is organized as follows. The main framework of the model is described in detail in Section 2. The relation between impact factors and certain elements is analyzed in Section 3. Section 4 endeavors to simulate the impact factors of four journals in distinct disciplines, and then compares the model data with the real data from 2005 to 2015 of those journals, intuitively. Finally, this paper is concluded in Section 5.

## 2. Model Formulation

The simulation model is discrete-timed and the unit of time is month. The main procedure of the model can be divided into two stages.

The first stage is the initial setup of the model, in which some essential information are empirically set, including:
1) The number of journals.
2) The number of issues of a journal published per year.
3) The number of articles published per issue.
4) The average review cycle. It should be explained that the average review cycle in the model is mainly determined by two time points, which are the time of an article being submitted and the time of the article being published.
5) The average number of references per article in a journal.
6) Assigning some relevant parameters.

The second stage is the design of citation behavior, which is the most important stage. In this study, inspired by a number of recent mathematical models on bibliometrics [18-20], we hypothesize that probability of an article to be cited by another article in the same discipline is jointly determined by three factors.
1) The intrinsic quality of an article $Q$.

In reality, owing to the fact that the quality of most of articles is medium leveled, those articles with extremely high or low quality would rarely appear in SCI journals. We thereby make a reasonable assumption that the intrinsic quality of the article can be scored as 1-10 points, with 10 being the best and 1 being the worst, and the overall quality of all articles follows the skewed distribution with certain expectation and variance [21], as illustrated in Fig.1.

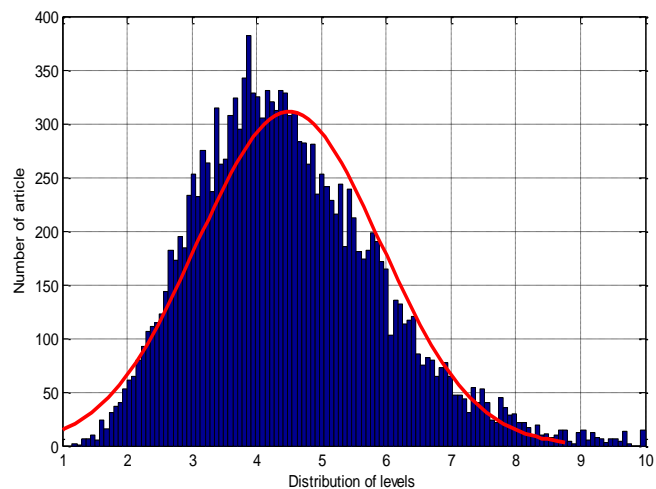

**Fig. 1.** Level distribution (1-10) of all articles with an expectation of 4.5 and a variance of 2. The number of journals is 10; the number of issues of a journal published per year is 12; the number of articles published per issue is 10; the time period is 13 years; and the total number of all articles is $10\times12\times10\times13=15600$.

2) The number of cites an article has.

Here we assume that the probability of an article to be cited is higher if the current number of citations of this article is greater [22]. The effect of the number of times an article has already been cited to the probability of the article to be cited in our model is depicted by a function $y=f(x)$, where $y$ is a multiplying factor to the ultimate probability, which follows several qualitative principles:

(i) The function is increasing in the interval $(0,+\infty)$.

(ii) The slope of the function is always decreasing in the interval $(0,+\infty)$.

(iii) $\lim_{x\to+\infty} f(x)=1$.

According to the above principles, we set the function as the following form.

$$y = \tanh(\frac{x}{\gamma}) \quad (1)$$

with $\tanh(\bullet)$ being the hyperbolic tangent function, $x\in N$ being the number of times the article has been cited, $y\in(0,1)$ being the probability of the article to be cited, and $\gamma\in R^+$ the parameter shaping the overall slope of the curve. Fig. 2 illustrates two curves of $y=f(x)$ with different parameter values.

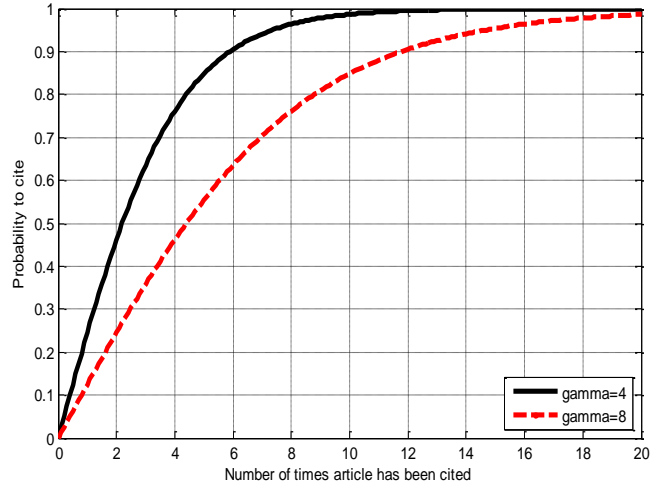

**Fig. 2.** Relation between effect to probability of an article to be cited and number of times the article has been cited.

Surely, one should be aware of the fact that the number of citations of an article is not the only criterion to measure the quality of an article.

**Remark 1.** The real purpose of choosing this hyperbolic function is to qualitatively analyze differences in citation behavior among different disciplines, rather than to accurately reproduce actual citation situation.

**Remark 2.** The greater the parameter $\gamma$, namely the less steep the curve, an article would be more inclined to be cited in the specific discipline.

3) The article age.

Due to the phenomena that there are obvious differences of timeliness of the research achievements between different basic sciences [23], for instance, in certain theoretical disciplines those relatively mature literatures that have been fully validated are more likely to be cited, whereas in contrast, scholars in several experimental disciplines prefer to cite some newer scientific achievements, in such a context we also find a function $y = g(x)$ for describing the relation between the effect to probability of an article to be cited by another article in the same discipline and the article age, which is the number of months between the publications of the cited and citing article. For rationality, this function should follow two qualitative principles.

(i) $\lim_{x \to -\infty} g(x) = 0$, $\lim_{x \to 0} g(x) = 1$.

(ii) The function is increasing in the interval $(-\infty, 0)$.

In accordance with such principles, the function can be expressed as follows:

$$y = \frac{1}{2}\tanh(\frac{x+\alpha}{\beta}) + \frac{1}{2} \tag{2}$$

where tanh(•) is also a hyperbolic tangent function; $x \in Z^-$ indicates the article age, with its unit being a month; $y \in (0,1)$ is the probability of the article to be cited; $\alpha$ is the parameter reflecting the horizontal translation of the curve; and $\beta$ is the parameter shaping the overall slope.

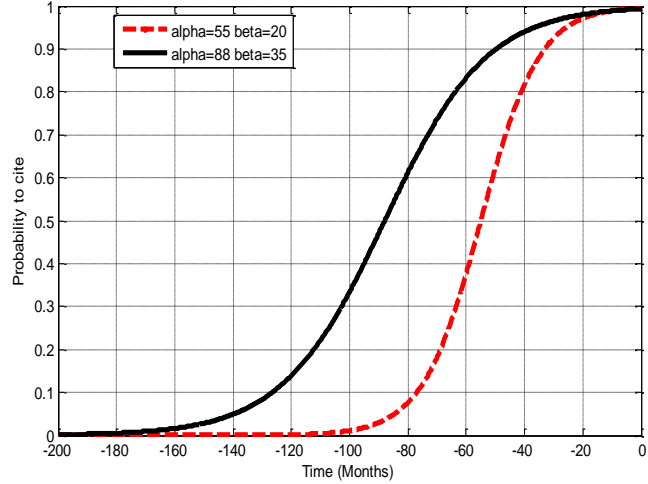

**Fig. 3.** Relation between probability of an article to be cited and article age.

**Remark 3.** The horizontal shift and the slope of the curve are determined by the parameters $\alpha$ and $\beta$, respectively. Qualitatively speaking, a flatter curve shows a longer citation life cycle of articles, while a steeper one indicates a reversed situation that these disciplines tend to cite newer articles.

Finally, according to the above design of citation behavior, the probability of an article to be cited in our model can be comprehensively defined as a combination of a series of factors, taking the following form.

$$P_{cite} = \frac{Q}{10} \cdot \tanh(\frac{N}{\gamma}) \cdot [\frac{1}{2}\tanh(\frac{T+\alpha}{\beta}) + \frac{1}{2}] \tag{3}$$

Note that in (3), $Q$ denotes the intrinsic quality of an article, $N$ denotes the number of cites an article has and $T$ denotes the article age.

Next we show how $P_{cite}$ is used to compute the impact factor. In this study, we develop a virtual citation program in which articles are published and cited in sequence. The articles are created one by one in our program, each time a new article is added, the database of the citation program is correspondingly incremented by one item, which can be deemed as a potential candidate for succedent citations. It should be noted that the probability of a new article to be cited is exactly determined by $P_{cite}$. As the bibliographic information of the created articles are available, according to the definition of impact factor, it can be calculated as follows:

$$IF_y = \frac{Cites_{y-1} + Cites_{y-2}}{Publications_{y-1} + Publications_{y-2}} \quad (4)$$

where *Publications* is the total number of articles a journal published in any given year $y-1$ and year $y-2$ and *Cites* is the number of cites by those articles in the year $y-1$ and year $y-2$.

Additionally, for simplicity, we uniformly set the model impact factor in the first two years to 1 for avoiding certain undesirable situations in the process of citation.

## 3. Relation Between Impact Factor and Certain Elements

In this section, we select four representative journals from the JCR database as the simulation reference, whilst the average impact factors of the four journals from 2005 to 2015, the average review cycles as well as the average numbers of references are collected and shown in Tab. 1. It should be explained that the data is obtained by manual counting. Specifically, we randomly select 300 article samples for each journal and calculate the average number of references. Similarly, according to the acceptation time and the publication time displayed in the articles, the average review cycles can be easily computed.

**Tab. 1** Integrated information of four actual journals

| No. | Journal Title | Average Impact Factors (2005-2015) | Average Review Cycles (Months) | Average Number of References |
|---|---|---|---|---|
| 1 | Nature Cell Biology | 19.20 | 5 | 63 |
| 2 | Nature Chemical Biology | 13.95 | 6 | 52 |
| 3 | IEEE Transactions on Automatic Control | 2.65 | 17 | 28 |
| 4 | Linear Algebra and its Applications | 0.96 | 9 | 18 |

By comparing and analyzing the data from Tab. 1, we speculate that the impact factor of a journal seems to be correlated with the average review cycle of the journal and the average number of references of the journal. To verify this speculation, we simulate the relation between the impact factor and certain particular elements of a journal. See Fig. 4 and Fig. 5.

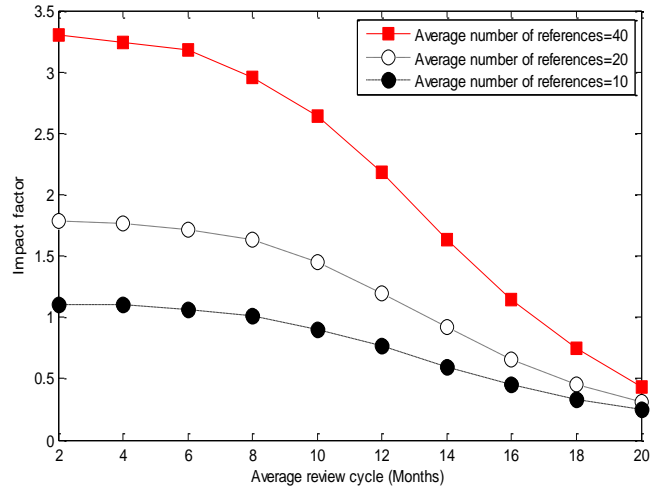

**Fig. 4.** Relation among impact factors, average number of references, and average review cycles with $\alpha=100$, $\beta=30$, and $\gamma=10$.

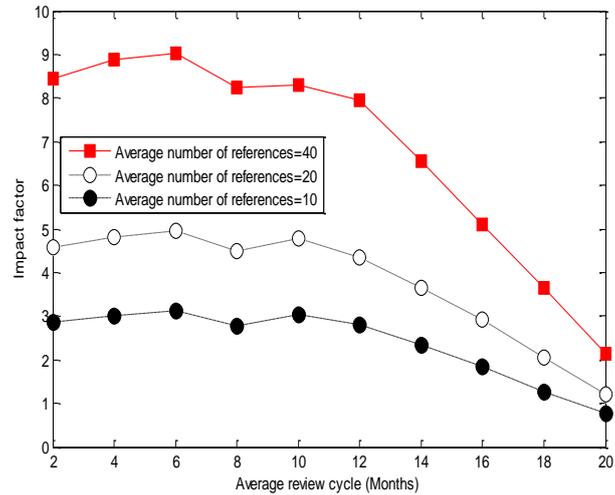

**Fig. 5.** Relation among impact factors, average number of references, and average review cycles with $\alpha=15$, $\beta=10$, and $\gamma=3$.

Observing the variation curves of Fig. 4 and Fig. 5, we find that the overall trends of impact factors are roughly analogous under different parameters ($\alpha=100, \beta=30$, & $\gamma=10$ and $\alpha=15$, $\beta=10$, & $\gamma=3$). It is evident that the impact factor of journals belonging to a specific discipline is intensively influenced by two factors, which are the average review cycle of and the average number of references in the routine. The impact factor of the journal with the average number of references 40 or 20 is inclined to be higher than the impact factor of another journal with the average number of references 10. This implicates that the greater the average number of references of a journal, the higher the impact factor of the journal would tend to be. Furthermore, we note that the impact factor of a journal is correspondingly decreasing with the average review cycle becoming longer (although with some small fluctuations), especially when the average review cycle reaches to 12 months.

Moreover, in the process of manual counting, by comparing the yearly distribution of references of these journals, we find that the yearly distribution of references of experimental disciplines (*Nature Cell Biology / Nature Chemical Biology*) are mainly concentrated in 2000 to 2015 (up to 98.5% and 86.4% respectively), and significantly in contrast, the percentages of references of journals in theoretical disciplines (*IEEE Transactions on Automatic Control / Linear Algebra and its Applications*) before 2000 are 29.2% and even 51.7%, respectively (Tab. 2). Such an observation indicates that the intuitive speculation about the monotonous correlation between JIF and the general range of yearly distribution of references conforms to the actual situation since the yearly distributions of references are explicitly distinguished among different basic sciences here.

**Tab. 2** Yearly distribution of references in four journals

| No. | Journal Title | Year | | |
|---|---|---|---|---|
| | | 2015-2010 (%) | 2009-2000 (%) | Before 2000 (%) |
| 1 | Nature Cell Biology | 57.1 | 41.4 | 1.5 |
| 2 | Nature Chemical Biology | 67.8 | 18.6 | 13.6 |
| 3 | IEEE Transactions on Automatic Control | 37.4 | 33.4 | 29.2 |
| 4 | Linear Algebra and its Applications | 12.8 | 35.5 | 51.7 |

Our distributed parallel model could provide an assistance for qualitatively and experimentally analyzing the relation of the impact factors and the yearly distribution of references, see Tab. 3 and Tab. 4.

**Tab. 3** Relation of impact factors and parameter values ( $\alpha$ & $\beta$ ) with average number of references being 30 and average review cycles being 4

| No. | $\alpha$ | $\beta$ | $\gamma$ | Average Impact Factors (2005-2015) |
|---|---|---|---|---|
| 1 | 90 | 40 | - | 2.2754 |
| 2 | 80 | 35 | - | 2.3207 |
| 3 | 70 | 30 | - | 2.4511 |
| 4 | 60 | 25 | - | 2.6753 |
| 5 | 50 | 20 | - | 3.0066 |
| 6 | 40 | 15 | - | 3.5657 |
| 7 | 30 | 15 | - | 4.2566 |
| 8 | 20 | 15 | - | 5.3283 |

**Tab. 4** Relation of impact factors and parameter values ($\alpha$ & $\beta$)
with average number of references being 20 and average review cycles being 10

| No. | $\alpha$ | $\beta$ | $\gamma$ | Average Impact Factors (2005-2015) |
|---|---|---|---|---|
| 1 | 90 | 40 | - | 1.2002 |
| 2 | 80 | 35 | - | 1.2479 |
| 3 | 70 | 30 | - | 1.3116 |
| 4 | 60 | 25 | - | 1.4572 |
| 5 | 50 | 20 | - | 1.6553 |
| 6 | 40 | 15 | - | 2.0608 |
| 7 | 30 | 15 | - | 2.6433 |
| 8 | 20 | 15 | - | 3.4259 |

Comparing the data in Tab. 3 and Tab. 4, under different conditions (average number of references is 30 & 20 and average review cycles is 4 & 10, respectively), we see a phenomenon that the average impact factors would increasingly grow with the decrease of parameter $\alpha$ and $\beta$. It is worth remarking that the parameter values of $\alpha$ and $\beta$ jointly reflect the citation life cycle of articles, where a steeper one, namely lesser $\beta$ indicates that the corresponding journals in some particular discipline tend to cite newer articles, while a greater $\beta$ signifies longer citation life cycle of articles and wider range of yearly distribution of references in that discipline.

## 4. Simulation Tests

Here according to the collected data (Tab. 1), we endeavor to simulate the trends of impact factors of four journals by adjusting appropriate parameters and initial settings in the model. After all, the primary aim of this section is only to test the reasonability of the model, rather than to reproduce the actual trends of impact factors from the proposed model. See Tab. 5 and Fig. 6.

**Tab. 5** Comparison of integrated information of actual journals and simulated journals

| No. | Journal Title | Average Impact Factors (2005-2015) | Average Review Cycles (Months) | Average Number of References |
|---|---|---|---|---|
| 1 | Nature Cell Biology | 19.20 | 5 | 63 |
| 2 | Journal 1 | 19.30 | 5 | 60 |
| 3 | Nature Chemical Biology | 13.95 | 6 | 52 |
| 4 | Journal 2 | 14.08 | 6 | 50 |
| 5 | IEEE Transactions on Automatic Control | 2.65 | 17 | 28 |
| 6 | Journal 3 | 2.65 | 17 | 30 |
| 7 | Linear Algebra and its Applications | 0.96 | 10 | 18 |
| 8 | Journal 4 | 0.94 | 10 | 20 |

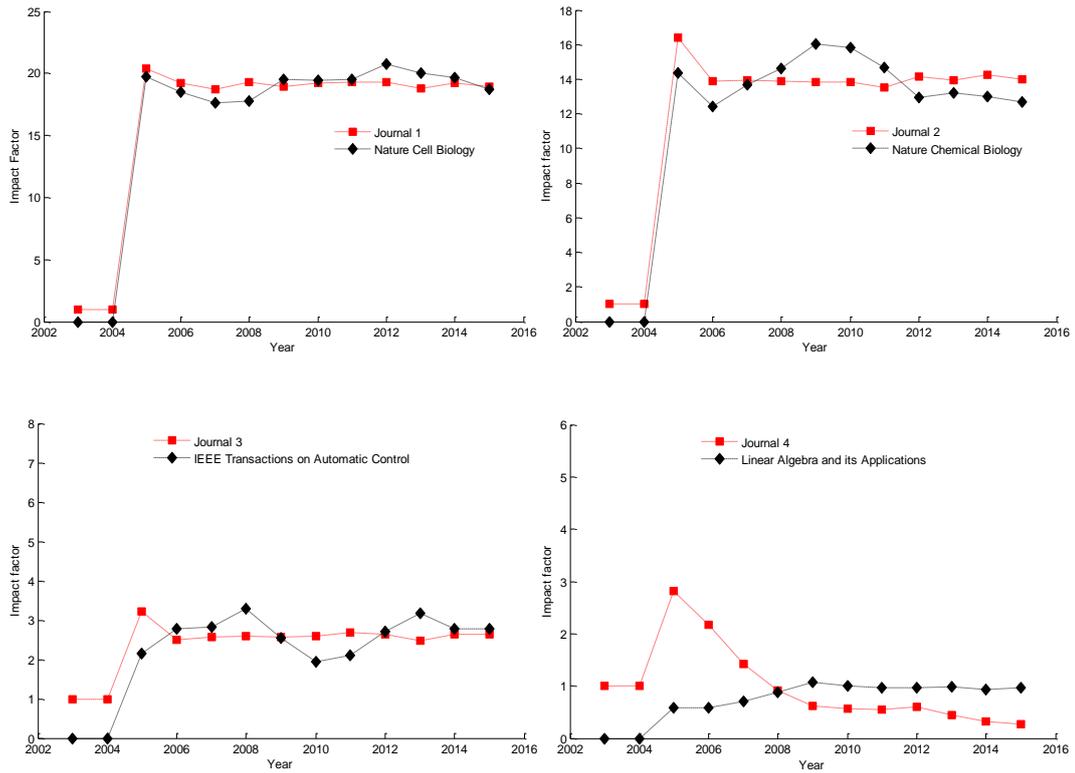

**Fig. 6.** Comparation of impact factors of four actual/simulated journals.

From the curves shown in Fig. 6, under certain conditions (the average review cycles, average number of references, and average impact factor are basically similar), we see that the trends of impact factors of different disciplines during a decade (2005-2015) can be approximately simulated by our experimental model. One also finds that there are significant differences of impact factors between different basic sciences, and the differences of impact factors between journals in experimental disciplines and theoretical disciplines could be as high as about 15. In this regard, although the quality of academic journals can be measured by the impact factor, it does not mean that impact factor is always proportional to the quality of journals, namely, the better the academic quality of a journal, the higher the impact factor of the journal certainly would be. Hence, it is unadvisable to indiscriminately compare the impact factors between journals of different disciplines when the universities and research institutions are measuring the performance of scientists.

## 5. Conclusions

The main purpose of this paper is to explore the principles of variation dynamics of impact factors among different disciplines. Our research attempts to provide a novel parallel systems based method that might help to understand the cause of certain phenomena such as the conspicuous difference of impact factor levels among different disciplines and further clarify the limitation and mechanism of impact factor for the process of academic evaluation. According to the simulation results and analysis on actual data, we infer that the overall impact factor level of journals in a discipline is intensively influenced by three factors, which are the average review cycle, the average number of references and the yearly distribution of references. Concretely speaking, a journal in a discipline which has a relatively shorter review cycle in the routine tends to have a higher impact factor. In addition, the typical number and the yearly distribution of references in a discipline not only present the timeliness of research achievements, but also contribute to affecting the overall impact factors. Objectively speaking, without the approach and the aid of simulation experiments here, it would be very difficult to derive the explicit evidence that implies the statistical correlations between the value of JIF and other elements in publication.

In fact, our attention to the variation of impact factors partially arises from a motivation to verify a speculation that blind comparation and indiscriminate application of impact factor might deteriorate the fairness of the academic evaluation. For example, impact factor is taken as an evaluation tool of journal quality, yet it is possible to be manipulated by the editors of journals, if with sufficient knowledge about its variation mechanisms.

The future work can be conducted sequentially along the current direction, combining approaches in scientometrics and social computing. As an instance, the current model could be extended by taking the manuscript submission behaviors into consideration; also, another possible work is to establish a dynamic index for more precisely measuring the academic quality of a journal based on our models.

## Acknowledgments

This work is supported by National Natural Science Foundation (NNSF) of China (Grants 61374054 & 61263002), by Fundamental Research Funds for the Central Universities (Grants 31920160003 & 31920170141), by Program for Young Talents of State Ethnic Affairs Commission (SEAC) of China (Grant 2013-3-21), and by Scientific Research Innovation Subject of Northwest Minzu University (Grant

yxm2015216).

## Competing Interests

The authors declare that they have no competing interests regarding the publication of this paper.

## References

[1] E. Garfield, "The history and meaning of the journal impact factor", *JAMA-J Am Med Assoc*, vol. 295, pp. 90-93, 2006.

[2] W. Glänzel and H. F. Mode, "Journal impact measures in bibliometric research", *Scientometrics*, vol. 53, pp. 171-193, 2002.

[3] N. Sombatsompop, T. Markpin and N. Premkamolnetr, "A modified method for calculating the Impact Factors of journals in ISI Journal Citation Reports: Polymer Science Category in 1997-2001", *Scientometrics*, vol. 60, pp. 217-235, 2004.

[4] W. Kuo and J. Rupe, "R-impact factor: Reliability-based citation impact factor", *IEEE Trans Reliab*, vol. 56, pp. 366-367, 2007.

[5] M. Gagolewski, "Scientific impact assessment cannot be fair", *J Informetr*, vol. 7, pp.792-802, 2013.

[6] G. Hirst, "Discipline Impact Factor: a method for determining core journal lists", *J Am Soc Inform Sci*, vol. 29, pp. 171-172, 1978.

[7] A. Hemmingsson, "Manipulation of impact factors by editors of scientific journals", *Am J Roentgenol*, vol. 178, pp. 767, 2002.

[8] Y. Guang, H.-Y. Dong and L. Wang, "Reliability-based citation impact factor and the manipulation of impact factor", *Scientometrics*, vol. 83, pp. 259-270, 2010.

[9] D.-H. Yang, X. Li, X.-X. Sun and J. Wan, "Detecting impact factor manipulation with data mining techniques", *Scientometrics*, vol. 109, pp. 1989-2005, 2016.

[10] E. Garfield, "Long-term vs. short-term journal impact: Does it matter?", *The Scientist*, vol. 12, pp. 11-12, 1998.

[11] P. Heneberg, "From excessive journal self-cites to citation stacking: Analysis of journal self-citation kinetics in search for journals, which boost their scientometric indicators", *PLoS ONE*, vol. 11, 2016.

[12] A. Haghdoost, M. Zare and A. Bazrafshan, "How variable are the journal impact measures?", *Online Inform Rev*, vol. 38, pp. 723-737, 2014.

[13] N. Cai, J.-W. Cao, H.-Y. Ma, and C.-X. Wang, "Swarm stability analysis of nonlinear

dynamical multi-agent systems via relative Lyapunov function", *Arab J Sci Eng*, 2014, vol. 39, pp. 2427-2434, 2014.

[14] N. Cai, C. Diao, and M. J. Khan, "A novel clustering method based on quasi-consensus motions of dynamical multi-agent systems", *Complexity*, 4978613, 2017.

[15] J.-X. Xi, M. He, H. Liu, and J.-F. Zheng, "Admissible output consensualization control for singular multi-agent systems with time delays", *J Franklin Inst*, vol. 353, pp. 4074-4090, 2016.

[16] F.-Y. Wang, "Back to the future: Surrogates, mirror worlds, and parallel universes", *IEEE Intelli Syst*, vol. 26, pp. 2-4, 2011.

[17] F.-Y. Wang, "Toward a paradigm shift in social computing: The ACP approach", *IEEE Intelli Syst*, vol. 22, pp. 65-67, 2007.

[18] M. Bras-Amorós, J. Domingo-Ferrer and V. Torra, "A bibliometric index based on the collaboration distance between cited and citing authors", *J Informetr*, vol. 5, pp. 248–264, 2011.

[19] G. V. Ionescu and B. Chopard, "An agent-based model for the bibliometric h-index", *Eur Phys J B*, vol. 86, pp. 426, 2013.

[20] I. U. Park, M. W. Peacey and M. R. Munafo, "Modeling the effects of subjective and objective decision making in scientific peer review", *Nature*, vol. 506, pp. 93-96, 2014.

[21] Daniel M. Herron, "Is expert peer review obsolete? A model suggests that post-publication reader review may exceed the accuracy of traditional peer review", *Surg Endosc*, vol. 26, pp. 2275-2280, 2012.

[22] Y.-T. Sun and B.-S. Xia, "The scholarly communication of economic knowledge: A citation analysis of Google Scholar", *Scientometrics*, vol. 109, pp. 1965-1978, 2016.

[23] L. Egghe and R. Rousseau, "The influence of publication delays on the observed aging distribution of scientific literature", *J Am Soc Inform Sci*, vol. 51, pp. 158-165, 2000.

# Appendix (Matlab Code for Review)

```matlab
clc
clear all;
close all;
num_of_journals = 10; % The number of journals
num_of_papers_per_journal = 120; % The number of papers per journal
average_issues_per_year = 12; % The number of issues of a journal pubilshed per year
average_ref_per_paper = 30; % The average number of references in a journal
average_review_cycle = 4; % The average review cycle
cur_article = 1;
alpha = 80;
beta = 60;
gamma = 36;
delta = 10;
quality_of_paper = gamrnd(10,0.45,15600,1); % The overall quality of all papers followed the skewed distribution with an expectation of 4.5
quality_of_paper = quality_of_paper-mod(quality_of_paper,1);
article = zeros(15600,105);
for i = 1:num_of_journals
    for j=1:13
        impact_factors(i,j) = 1;
    end
end

month=1;
for year=1:13
    for issue=1:average_issues_per_year
        for i=1:num_of_journals
            for j=1:10    %number of papers per issue
                article(cur_article, 1) = quality_of_paper(randi(length(quality_of_paper))); % The intrinsic quality of a paper
                if article(cur_article, 1)>10
                    article(cur_article, 1) = 10;
                end
                    if article(cur_article, 1)<1
                    article(cur_article, 1) = 1;
                end
                article(cur_article, 2) = month; % article age
                article(cur_article, 3) = i; % published journal
                article(cur_article, 4) = rand*average_ref_per_paper*2;
                article(cur_article, 4) = article(cur_article, 4)-mod(article(cur_article, 4),1); % The number of references in a paper.
                if article(cur_article, 4)<10
                    article(cur_article, 4) = 10;
```

```matlab
            end
            article(cur_article, 5) = 0; % The number of times the paper has been cited.
            cur_ref=1;
            if month>24 % We assume that the citing begins after 1 year.
                while cur_ref<=article(cur_article, 4)
                    dice=rand*cur_article;
                    candidate_ref=dice-mod(dice,1);
                    if candidate_ref<1
                        candidate_ref = 1;
                    end
                    if month-article(candidate_ref,2)>average_review_cycle % if the time is matched, please continue.
                        probability_to_cite = article(candidate_ref,1)*(0.5*tanh((article(candidate_ref, 2)-month+alpha)/beta)+0.5)*tanh((article(candidate_ref, 5)+delta)/gamma);
                        % The probability of a paper to be cited is jointly determined by the three factor (1. intrinsic quality 2. number of times 3. paper age)
                        dice = rand;
                        if dice<probability_to_cite % Should be cited
                            article(cur_article, 5+cur_ref) = candidate_ref;
                            article(candidate_ref, 5) = article(candidate_ref, 5)+1;
                            % The number of references, each a new paper is added, this value is incremented by 1.
                            ref_year = article(candidate_ref,2)/12-mod(article(candidate_ref,2)/12,1)+1;
                            ref_journal = article(candidate_ref, 3);
                            if month>24 && ref_year<year && ref_year>year-2 % Compute the Impact Factor
                                impact_factors(ref_journal, year)=impact_factors(ref_journal, year)+1;
                            end
                            cur_ref = cur_ref+1;
                        end
                    end
                end
            end
            cur_article = cur_article+1;
        end
    end
    month=month+1;
end
if year>2
    for i=1:num_of_journals
        impact_factors(i,year)=impact_factors(i,year)/(2*num_of_papers_per_journal); % Compute the Impact Factor
    end
end
end
```